\documentclass[prd,showpacs,letterpaper,twocolumn]{revtex4}%
\usepackage{graphicx}
\usepackage{bm}
\usepackage{epsf}
\usepackage{rotating}
\usepackage{epsfig,graphics,rotate,color}
\usepackage{wrapfig}
\usepackage{amssymb}
\usepackage{amsmath}
\usepackage{amsfonts}
\usepackage{array,hhline,dcolumn}%
\usepackage{multirow}
\providecommand{\U}[1]{\protect\rule{.1in}{.1in}}
\begin{document}
\title{Precision $\bar \nu_e$~--~electron Scattering Measurements with
  IsoDAR to Search for New Physics}
\author{ J.M. Conrad$^{1}$, M.H. Shaevitz$^{2}$, I. Shimizu$^3$
  J. Spitz$^1$, M. Toups$^{1}$\footnote{Corresponding author. Matt Toups, 
        E-mail address: mtoups@mit.edu, 
        postal address: Fermilab, MS309, POBox 500, Batavia, IL, 60510, USA, 
        phone number: +1-630-840-4759, fax number: +1-630-840-6520}, and L. Winslow$^{4}$}
\affiliation{$^{1}$ Massachusetts Institute of Technology, Cambridge, MA 02139, USA}
\affiliation{$^{2}$ Columbia University, New York, NY 10027, USA}
\affiliation{$^{3}$ Tohoku University, Sendai, 980-8578, Japan}
\affiliation{$^{4}$ University of California, Los Angeles, Los Angeles, California 90095, USA}

\begin{abstract}

IsoDAR provides a pure and intense $\bar \nu_e$ source with an average energy of 6.5~MeV produced through
$^8$Li $\beta$-decay.  This source can be paired with a large scintillator detector, such as KamLAND, to produce
a sample of $\bar \nu_e$-electron scatters that is more than five
times larger than what has been collected before.  Such a sample would allow for a 3.2\% measurement of $\sin^2\theta_W$ and a sensitive search for non-standard interactions. 
  
\end{abstract}

\maketitle

\section{Introduction}

A large sample of antineutrino-electron scattering (ES) events ($\bar \nu_e
+ e^- \rightarrow \bar \nu_e + e^-$) allows for sensitive searches for
Beyond Standard Model physics.  In the Standard Model, the ES cross
section depends only on kinematic terms and the weak couplings,  $g_V$
and $g_A$, or, equivalently, $\sin^2 \theta_W$.  There are no complications arising from strong interaction as in neutrino-quark scattering, because ES is purely leptonic.  Currently, 
$\sin^2 \theta_W$ is well known from measurements outside of the 
neutrino sector \cite{EWWG}, and the {\it ab initio} prediction for this two-lepton scattering
process is therefore very precise.  However, a rich variety of new physics in the neutrino sector can affect the ES cross section. Such physics can include heavy partners which mix with the light neutrinos or new $Z^\prime$s that couple only to neutrinos~\cite{AndredeG}.  Recent work~\cite{Valle} has investigated the largely complementary sensitivity of the neutrino-electron scattering ($\nu_e + e^- \rightarrow \nu_e + e^-$) cross section to these effects in the context of the proposed LENA detector~\cite{Wurm}.

In this paper, we outline a precision study using the proposed electron antineutrino source, IsoDAR~\cite{PRL} which is being developed as part of the DAE$\delta$ALUS program~\cite{EOI}.
The high event rate provided by this low energy source leads to 
the possibility of precision measurements of the couplings ($g_V$
and $g_A$) and $\sin^2 \theta_W$. Along with these analyses, we explore IsoDAR's sensitivity to  nonstandard interactions (NSIs)--new physics introduced into the theory via an
effective 4-fermion term in the Lagrangian~\cite{Berezhiani:2001rs}.
NSIs induce corrections to the Standard Model couplings, $g_V$ and $g_A$.
An observed deviation from the Standard Model expectation, indicative of new physics,
could dramatically change our evolving understanding of neutrino properties and interactions.   

\section{The IsoDAR source}

The IsoDAR antineutrino source~\cite{PRL}, when combined with the KamLAND
detector~\cite{kamMix}, can collect more than 2400~ES events in a five
year run. This estimate is smaller than that reported in Ref.~\cite{PRL} as a number of analysis cuts have been introduced.  Such a collection of $\bar\nu_e$ ES events would be the largest to date and can
be compared to the samples from the Irvine experiment 
(458 events from 1.5 to 3 MeV~\cite{Irvine:1976}); 
TEXONO (414~events from 3 to 8~MeV~\cite{TEXONO:2012}); 
Rovno (41 ~events from 0.6 to 2~MeV~\cite{ROVNO:1993}); and 
MUNU (68~events from 0.7 to 2~MeV~\cite{MUNU:2005}).  

IsoDAR~\cite{PRL} is a cyclotron that will accelerate protons to 60~MeV. The protons impinge on a $^9$Be target to produce an abundant source of neutrons. The neutrons subsequently enter a surrounding 99.99\% isotopically pure $^7$Li sleeve, where
neutron capture results in the creation of $^8$Li. This unstable isotope undergoes $\beta$ decay 
to produce an isotropic $\bar \nu_e$ flux with an average energy of
$\sim$6.5~MeV and an endpoint of $\sim$13 MeV. The $\bar \nu_e$ interact in the scintillator detector via ES and inverse beta decay (IBD), $\bar \nu_e +p \rightarrow e^+ + n$.  
Along with being the signal channel for the sterile neutrino search described in Ref.~\cite{PRL}, the latter interaction is important for an ES measurement as it provides a method to constrain the normalization 
of the flux, as described in Ref.~\cite{ConradLinkShaevitz}.  
We note, however, that the misidentification of IBD events as $\bar \nu_e$ events represents a significant source of background. The key experimental parameters are summarized in Table~\ref{assumptions}.

\begin{table}[tbh]
  \begin{center}
    {\footnotesize
      \begin{tabular}{|c|c|} \hline
        Accelerator  & 60~MeV/amu of H$_2^+$  \\  \hline
        Power  & 600~kW  \\  \hline
        Duty cycle  & 90\%  \\  \hline
        Run period  & 5~yrs (4.5~yrs live)  \\  \hline
        Target, sleeve   & $^9$Be, $^7$Li (99.99\%) \\  \hline
        $\overline{\nu}_e$ source  & $^8$Li $\beta$ decay \\   \hline
        $\overline{\nu}_e$ $\langle E_\nu\rangle$ & 6.4~MeV  \\  \hline
        $\overline{\nu}_e$ flux  & 1.29$\times10^{23}$
$\overline{\nu}_e$ \\  \hline \hline
        Detector  & KamLAND   \\  \hline
        Fiducial mass  & 897 tons   \\  \hline
        Target face to detector center  & 16.1~m   \\  \hline
      \end{tabular}
      \caption{ The IsoDAR experiment's main characteristics, as presented in
          Ref.~\cite{PRL}}.\label{assumptions}}
\end{center}
\end{table}

\section{$\bar  \nu_e$-electron elastic scattering \label{es_formalism}}

The neutral current and charged current both contribute to the ES cross section.  
The ES Standard Model differential cross section is given by:
\begin{align}
\frac{d\sigma}{dT} = \frac{2 G_F^2 m_e}{\pi}
\big[ 
g^2_R + g^2_L(1 - \frac{T}{E_\nu})^2 - g_R g_L \frac{m_e T}{E^2_\nu}
\big]~,
\label{glgrxs}
\end{align}

where $g_R= {{1}\over{2}}(g_V-g_A)$, $g_L={{1}\over{2}}(g_V+g_A)$, $E_{\nu}$ is the incident $\overline{\nu}_{e}$ energy, $T$ is
the electron recoil kinetic energy, $G_F$ is the Fermi coupling constant, and $m_e$ is the mass of the electron.  The coupling constants at tree
level can be expressed as:
\begin{equation}
g_L={{1}\over{2}}+\sin^2\theta_W, ~~g_R=\sin^2\theta_W.
\end{equation}
Therefore, allowed ranges for $g_V$ and $g_A$ as well as $\sin^2\theta_W$ can be extracted from a measurement of the differential ES cross section.

On the other hand, the weak
mixing parameter, $\sin^2 \theta_W$, is related to $G_F$, $M_Z$ and $\alpha$ by
$\sin^2 2\theta_W=(4 \pi \alpha)/(\sqrt{2} G_F M_Z^2)$. Precision measurements 
at colliders~\cite{Erler} and from muon decay~\cite{mulan}
therefore lead to an absolute prediction for the ES cross section at
tree level.  Thus, given a precise prediction for the ES process, we can look for beyond Standard
Model physics effects that can cause a deviation from expectation in the measured cross
section. 

NSI terms modify the cross section for ES through apparent changes to the measured couplings in the following way:
\begin{eqnarray}\label{epsilonxsec}
\hspace{-1cm}
\frac{d\sigma(E_{\nu}, T)}{dT} &=& \frac{2 G_F^2 m_e}{\pi}[ (\tilde 
g_R^2+\sum_{\alpha \neq e}
|\epsilon_{\alpha e}^{e R}|^2) + \nonumber \\
& &(\tilde g_L^2+\sum_{\alpha \neq e}
|\epsilon_{\alpha e}^{e L}|^2)\left(1-{T \over E_{\nu}}\right)^2 -
\nonumber\\ 
& &(\tilde g_R \tilde g_L+ \sum_{\alpha \neq e}|\epsilon_{\alpha e}^{e R}||
\epsilon_{\alpha e}^{e L}|)m_e {T \over E^2_{\nu}}]~,
\end{eqnarray}
where $\tilde g_R= g_R+\epsilon_{e e}^{e R}$ and $\tilde g_L=g_L+\epsilon_{e e}^{e L}$.  The corresponding cross section for neutrinos can be obtained by interchanging $\tilde g_L$ with $\tilde g_R$  in Eq.~\ref{epsilonxsec}. The NSI parameters are $\epsilon_{e\mu}^{e LR}$
and $\epsilon_{e\tau}^{e LR}$, which are associated with flavor-changing-neutral currents, and $\epsilon_{ee}^{e
  LR}$, which are called non-universal parameters. As the former are well
constrained for muon flavor~\cite{Davidson:2003ha} and lepton flavor
violating processes are strongly limited in general, we neglect these
when considering IsoDAR's sensitivity to NSI. That is, we focus on the
two relevant non-universal parameters $\epsilon_{e e}^{eLR}$ and set the four others to
zero. This is also a matter of simplicity and convenience, given the
complications that can arise when making assumptions about multiple
terms that have the potential to cancel each other. We note that given
some set of assumptions, sensitivity to the poorly constrained
parameters $\epsilon_{e\tau}^{e L}$ and $\epsilon_{e\tau}^{e R}$ 
is also available.

A precision measurement of the ES cross section requires an experiment
which has excellent reconstruction capabilities, a precise
understanding of the flux normalization,  reasonably low
backgrounds that are well constrained by direct measurement, and substantial statistics.
The approach described here follows the proposed analysis of 
Ref.~\cite{ConradLinkShaevitz}, which examined an ES
cross section measurement at a reactor-based antineutrino source. The IsoDAR analysis has a considerable advantage over reactor-based measurements because the $^8$Li-induced flux peaks well above 3~MeV, where environmental backgrounds are substantially decreased. Furthermore, beam-off periods, which can be rare for commercial reactor sources, allow a determination of non-beam-related backgrounds in the case of IsoDAR.

\section {Signal and Background} \label{sec:signal_and_background}
Antineutrino-electron scattering events are simply characterized by the outgoing electron's energy in scintillation-based detectors. However, directly evaluating Eq.~\ref{glgrxs} requires the reconstruction of both $T$
and $E_\nu$.  The electron recoil kinetic energy, $T$, is equivalent to 
the visible energy in the KamLAND detector, $E_{\mathrm{vis}}$.  Unfortunately, $E_\nu$  cannot be reconstructed in KamLAND because the exiting antineutrino carries away an undetectable amount of energy and the outgoing electron's angle cannot be resolved.   As a result, our analysis strategy is to consider events in bins of $E_{\mathrm{vis}}$, and to integrate over all $E_\nu$ values
that can contribute to these populations. 

The uncertainty on the ES prediction is dominated by the normalization uncertainty on the antineutrino flux from the IsoDAR source. Following the method of Ref.~\cite{ConradLinkShaevitz}, this normalization will be determined from the observed IBD events that can be well isolated using the delayed coincidence of the prompt positron signal and delayed $2.2~$MeV neutron capture signal.  The precision of this method is limited by the 0.7\% uncertainty on the KamLAND IBD efficiency~\cite{kamMix}, which dominates over the 0.2\% IBD cross section error and the 0.1\% statistical error, given the nominal 5~year IsoDAR run expected.  

\begin{figure}[tb]
	\begin{center}
		\includegraphics[width=0.49\textwidth]{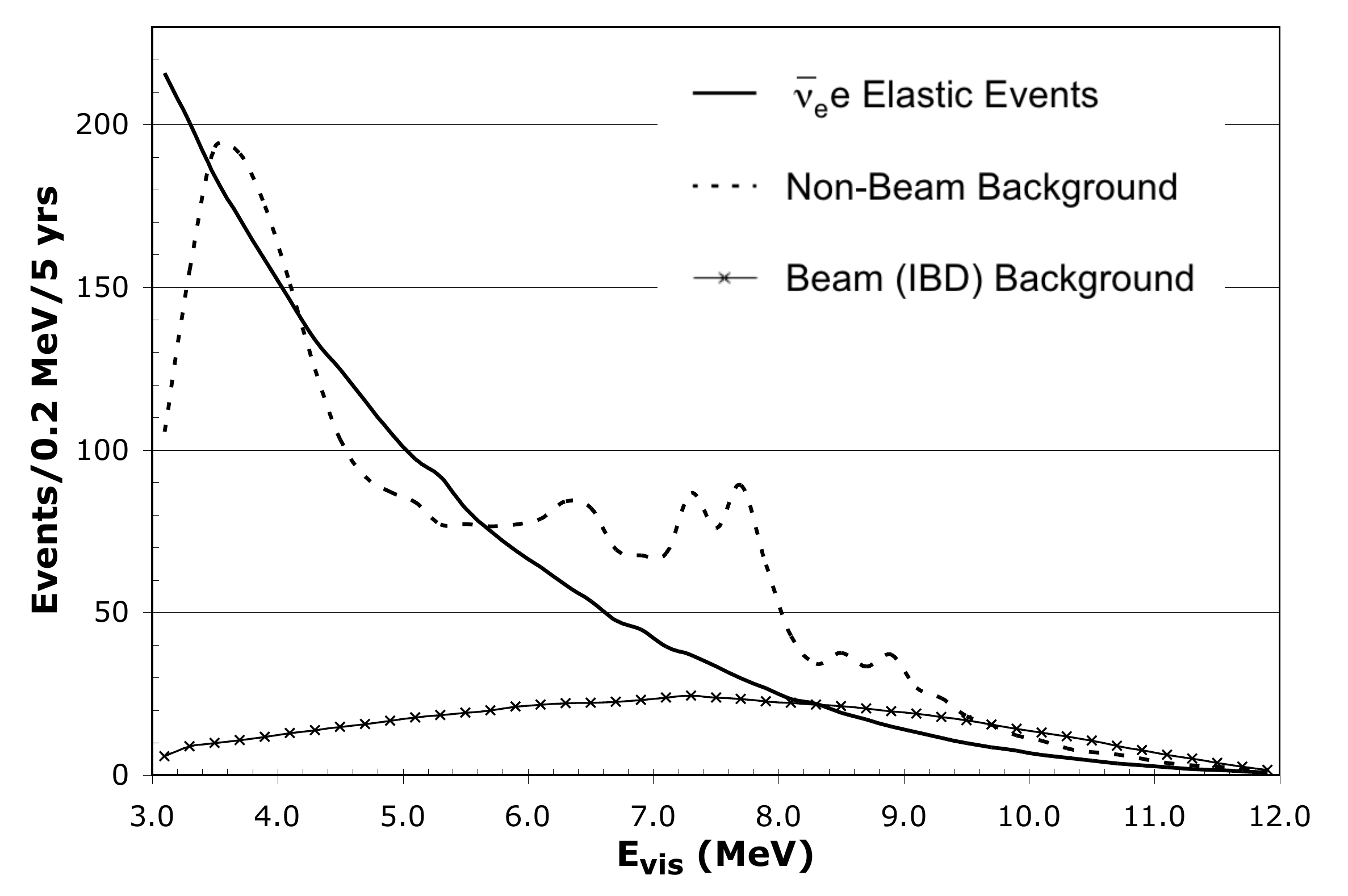}
	\end{center}
	\caption{\label{events_vs_evis} The number of signal and background events as a function of visible energy.
	The thick solid line shows the ES signal events, the dashed line shows the non-beam background, and the
	thin solid line with x's shows the misidentified IBD beam background. The distributions include an energy smearing of $\delta E_{\mathrm{vis}} = 0.065\cdot\sqrt{E_{\mathrm{vis}}\mathrm{(MeV)}}$.}
\end{figure}

A series of cuts are applied to mitigate ES backgrounds.  To reduce backgrounds from the decay of light isotopes produced by cosmic muon spallation in the detector, we employ the KamLAND muon veto cuts from Ref.~\cite{kamBoron}.  For muons with poorly reconstructed tracks and those with unusually high light levels, a 5~s veto is applied throughout the detector ($\Delta T_\mu>5~$s).  However, for well-tracked muons that do not have unusually high light levels, the $\Delta T_\mu>5~$s veto is applied in a 3~m radius ($\Delta R_\mu>3~$m) around the muon track and a $\Delta T_\mu>200~$ms veto is applied throughout the remainder of the detector.  The muon veto results in a dead time of $37.6\pm0.1\%$.  To separate the ES signal from low energy backgrounds, we use a low visible energy cut of $E_{\mathrm{vis}}>3$ MeV.  To remove backgrounds from external sources of radioactivity, a radial cut of $R<5.0$ m is applied to the reconstructed event vertex.  Finally, to reduce the positron background from IBD interactions, candidate ES events are rejected if there is a subsequent delayed event satisfying $E^d_{\mathrm{vis}}>1.8$~MeV, $R_d<6.0~$m, and $\Delta T_d<2~$ms, where $E^d_{\mathrm{vis}}$ is the visible energy of the delayed event, $R_d<6.0~$m is the reconstructed radial position of the delayed event vertex, and $\Delta T_d$ is the elapsed time to the delayed event.  The rate of triggers with $E>1.8~$MeV and $R<6.0~$m in KamLAND is 0.65 Hz, implying an IBD veto dead time of 0.1\%, which we neglect in this analysis. Table~\ref{cuts} summarizes all the cuts applied to reduce ES backgrounds. Figure~\ref{events_vs_evis} shows the ES and background events as a function of $E_{\mathrm{vis}}$ for a five year run with the
parameters given in Table~\ref{assumptions}. Table~\ref{events} shows the total number of events expected with  3~MeV~$<E_{\mathrm{vis}}<$~12~MeV. 

\begin{table}[tb]
\begin{center}
\begin{tabular}{lr}
\hline
\multicolumn{2}{c}{\textbf{Muon Veto}}\\
\hline \hline
All muons & $\Delta T_\mu>200~$ms\\
\hline
Well-tracked muons  & $\Delta T_\mu>5~$s for $\Delta R_\mu<3~$m \\
\hline
Poorly-tracked muons & $\Delta T_\mu>5~$s \\
\hline
\multicolumn{2}{c}{\textbf{ES Selection Cuts}}\\
\hline \hline
\multicolumn{2}{c}{$E_{\mathrm{vis}}>3$ MeV}\\
\multicolumn{2}{c}{$R<5.0$ m}\\
\hline 
  \multicolumn{2}{c}{\textbf{IBD Veto}} \\
  \hline \hline
Events with $E^d_{\mathrm{vis}}>1.8~$MeV~~ & ~~$\Delta T_d>2~$ms for $R_d<6.0~$m\\
\hline

\end{tabular}     
\end{center}
\caption{ {Summary of cuts used to reduce the ES backgrounds. The symbols are defined in the text.  The phrase ``poorly-tracked muons'' above also refers to muons which produce unusually high light levels. Further details can be found in the text. } \label{cuts}}
\end{table}

In the following subsections we provide more information on the calculation of the expected background rate and
$E_{\mathrm{vis}}$ dependence (i.e. ``shape'').  The backgrounds can be grouped into
beam-related backgrounds, which are dominated by IBD events, and
non-beam backgrounds, arising from solar neutrino interactions,
muon spallation, and environmental sources.  We envision a data-driven background estimation strategy, in which the non-beam backgrounds are measured in KamLAND data collected prior to the realization of the IsoDAR source and beam-related IBD backgrounds are estimated in KamLAND data after the IsoDAR source turns on.  Table~\ref{backevents} provides a summary of the non-beam backgrounds from 3--12 MeV before energy smearing is taken into account.

\begin{table}[tb]
\begin{center}
\begin{tabular}{c|c}
\hline
           &     Events \\
\hline
   Elastic scattering (ES) &       2583.5 \\

IBD Mis-ID Bkgnd &        705.3 \\

Non-beam Bkgnd &       2870.0 \\
\hline
Total   &  6158.8  \\
\hline
\end{tabular}     
\end{center}
\caption{ {Total signal and background events in KamLAND with $E_{\mathrm{vis}}$ between 3--12 MeV including an energy smearing of $\delta E_{\mathrm{vis}} = 0.065\cdot\sqrt{E_{\mathrm{vis}}\mathrm{(MeV)}}$ given the IsoDAR assumptions in Table~\ref{assumptions} and the cuts listed in Table~\ref{cuts}.} \label{events}}
\end{table}

\begin{table}[tb]

\begin{center}
\begin{tabular}{c|c}
\hline
 & Events \\
\hline
$^{8}$B Solar Neutrino & 890.1 \\
$^{208}$Tl  &  594.3 \\
External $\gamma$ Stainless &   227.4 \\
External $\gamma$ Rock &  533.7 \\
Spallation $^8$B & 42.5 \\
Spallation $^8$Li & 94.9 \\
Spallation $^{11}$Be & 490.0 \\
\hline
Total   &  2872.9  \\
\hline
\end{tabular}     
\end{center}
\caption{ {Total non-beam background events in KamLAND in the visible energy range 3--12 MeV given the IsoDAR assumptions in Table~\ref{assumptions} and the cuts listed in Table~\ref{cuts}.} \label{backevents}}
\end{table}

\subsection{Misidentified IBD events from beam interactions}
The primary beam-on background is due to misidentified IBD events.
Notably, the beam-on background from both fast and thermal neutrons is negligible. The IsoDAR source is designed with shielding to slow fast neutrons and reduce this background to a negligible level.  For thermal neutrons that leak into the fiducial volume, the visible energy from capture on Hydrogen is below the 3~MeV cut. 

The IsoDAR source produces $8 \times  10^5$ IBD interactions
over five years in the 897~ton fiducial mass KamLAND detector.  Most of the IBD
interactions can be removed from the ES sample by rejecting any events 
that are followed by a delayed neutron capture on Hydrogen.  However, even if
just 1\% of these events leak into the ES sample, then the IBD contribution becomes
the single largest background in this analysis. The KamLAND IBD identification analysis has evolved in time from simple time- and space-based cuts~\cite{kamMix1, kamMix2} to a more sophisticated likelihood-based selection~\cite{kamMix3, kamBoron, kamMix}. These cuts are chosen to maximize the purity of the IBD sample, and have an efficiency of around $\sim$90\%, with the precise value depending on the analysis. In this analysis, we strive to maximize the IBD detection efficiency so as to reduce the misidentified IBD background. 

In order to eliminate IBD background events, we reject any events in the ES sample which have a subsequent delayed event satisfying $E^d_{\mathrm{vis}}>1.8$~MeV, $R_d<6.0~$m, and $\Delta T_d<2~$ms, where $E^d_{\mathrm{vis}}$ is the visible energy of the delayed event, $R_d<6.0~$m is the reconstructed radial position of the delayed event vertex, and $\Delta T_d$ is the elapsed time to the delayed event.  To estimate the $E^d_{\mathrm{vis}}$ cut rejection efficiency, we use an energy resolution of $\delta E_{\mathrm{vis}}/E_{\mathrm{vis}} = 6.5\%/\sqrt{E_{\mathrm{vis}}\mathrm{(MeV)}}$~\cite{kamMix}.  Since over 99.99\% of IBD delayed neutrons capture on either $^{1}$H (2.2 MeV $\gamma$) or $^{12}$C (4.95 MeV $\gamma$)~\cite{Kazumi}, the low energy $E^d_{\mathrm{vis}}$ cut introduces a negligible IBD rejection inefficiency.  We estimate the $R_d$ cut rejection efficiency with a toy Monte Carlo simulation incorporating $\gamma$-ray attenuation lengths computed for the KamLAND scintillator~\cite{Enomoto} and a realistic PDF for the IBD prompt--delayed event distance ($\Delta R$) distribution derived from KamLAND AmBe calibration source data~\cite{ODonnell}.  For IBD events generated in a $R < 5.0~$m fiducial volume, we find that the $R_d<6.0~$m cut rejection efficiency is 99.80\%.  Finally, we use a mean neutron capture time in the KamLAND target of $207.5 \pm 2.8~\mu$s~\cite{kamMix} to estimate the $\Delta T_d$ cut rejection efficiency.  We find that less than 0.01\% of neutron captures occur after 2 ms.  However, there is an additional IBD rejection inefficiency at small $\Delta T_d$ due to pile-up events.  This inefficiency can be avoided by using detailed PMT waveform data and hit time distributions.  Based on the $^{212}$Bi-$^{212}$Po rejection efficiency reported in~\cite{KamZen}, we assume an 80\% rejection efficiency for IBD events with $\Delta T_d<0.5~\mu$s, giving a total $\Delta T_d<2~$ms cut rejection efficiency of 99.95\%.  Based on these estimates, we compute a combined IBD rejection efficiency of 99.75\%.

This combined IBD rejection efficiency can be estimated from a full volume calibration of the KamLAND detector with an Am/Be neutron source. We assume a data sample of 50,000 Am/Be delayed neutron events will allow for a statistics-limited measurement of the combined $E^d_{\mathrm{vis}}>1.8$~MeV, $R_d<6.0~$m, and $\Delta T_d<2~$ms cut rejection efficiency, and therefore use an IBD rejection efficiency of $99.75\% \pm 00.02\%$.

\subsection{Solar neutrino background}
The neutrino-electron elastic scattering of $^{8}$B solar neutrinos is a
background to the ES measurement with
IsoDAR. Super-Kamiokande produces the most precise measurement of the oscillated $^{8}$B flux using neutrino-electron scattering. The current measurement is  2.32$\pm$0.04(stat)$\pm$0.05(syst) $\times~10^6$~cm$^{-2}$s$^{-1}$~\cite{skSolar}. This corresponds to
4.10$\pm$0.11~events per kiloton-day in KamLAND before the application
of the analysis threshold assuming the standard scintillator
composition and using the neutrino-electron scattering cross-section
with radiative corrections from Ref.~\cite{solarCrossSect}. This
number has some dependence on  $\sin^2\theta_W$ which is neglected in
this analysis. 

The spectral shape of this background is also included. It is calculated using the same cross section from
Ref.~\cite{solarCrossSect} and the neutrino spectrum from
Ref.~\cite{shape8B}. The effect of neutrino oscillation is included
using the Standard Solar Model AGS2009~\cite{ssm2009} and the
oscillation parameters from the KamLAND global analysis in 
Ref.~\cite{kamMix}.

\subsection{Spallation backgrounds}

High energy beta decays of light isotopes produced in muon spallation
are an important subset of beam-off backgrounds. The production of
these isotopes in KamLAND is studied in detail in
Ref.~\cite{kamSpall}. The muon veto cuts
described above are adopted to mitigate the effect of this
background. These eliminate all isotopes with half-lives shorter
than $\sim$1~s, leaving only three isotopes that contribute to this
background above the 3~MeV analysis threshold: the $\beta^{+}$ decay of $^{8}$B ($\tau$=1.11~s,
Q=18~MeV)~\cite{nucDataLiB}, the $\beta^{-}$ decay of $^{8}$Li
($\tau$=1.21~s, Q=16~MeV)~\cite{nucDataLiB}, and most importantly the
$\beta^{-}$ decay of $^{11}$Be ($\tau$=19.9~s, 
Q=11.5~MeV)~\cite{nucDataBe}.

Both the production rate and the spectral shapes of these three isotopes
are included. 
The production rates are summarized in Table~\ref{backevents}. 
In the final analysis, beam-off data will be used to determine the shape
and rates.  Here, we use the spectra from Refs.~\cite{shape8B,
  shape8B8Li} for 
$^{8}$B and $^{8}$Li, which are complicated by the decay through a wide
excited state of $^{8}$Be.  The $^{11}$Be decay is calculated using the
standard beta-decay shape, accounting for corrections due to
forbiddenness and the deposition of energy by de-excitation gammas or
heavier particles in branches to excited states of $^{11}$B. A simple energy response is assumed for
KamLAND with no scintillator quenching for gammas and
electrons/positrons and with a quenching factor of 10 assumed for
 alpha particles.

\subsection{External gamma ray background}

The external gamma ray background above 3~MeV of deposited energy due
to the rock and stainless steel detector vessel was calculated using a Geant4 Monte Carlo simulation~\cite{geant4} with simplified KamLAND geometry. The simulation was tuned to match KamLAND data~\cite{kamBoron}. The total external gamma ray background above 3~MeV in the KamLAND target volume is 18~events/day from the rock and 56~events/day from the stainless steel vessel.
We use the simulation to scale these background rates to a fiducial radius of
5.0~m and obtain 1.275 events/kton-day and 0.543 events/kton-day from the rock and
steel, respectively. The total number of external gamma ray background events
are summarized in Table~\ref{backevents}. 

\subsection{$^{238}$U and $^{232}$Th background}

The decays of the daughters of the $^{238}$U and $^{232}$Th chains
within the liquid scintillator can produce backgrounds to this
measurement. Above the 3~MeV analysis threshold, the only beta decay
is $^{208}$Tl (Q = 5.0~MeV,  $\tau_{1/2}$=3.05~min)~\cite{nucData208Tl}
from the $^{232}$Th chain. There are several alpha decays above 3~MeV;
however, due to scintillator quenching they reconstruct below the
analysis threshold. Assuming $^{232}$Th contamination levels similar
to the low-background phase of KamLAND of
$(1.12\pm0.21)\times10^{-17}~$g/g~\cite{keefer2009, kyohei} leads to
$R_{^{208}\mathrm{Tl}}=1.42 \pm 0.27$~events/kiloton-day before the application
of the analysis threshold. The spectral shape is calculated using the
standard beta-decay shape, accounting for corrections due to
forbidden decays and the deposition of energy by de-excitation gammas in
branches to excited states of $^{208}$Pb as was done for the $^{11}$Be
shape. Once again, a simple energy response model is used to model the
detector response.

\section{Analysis \label{analysis}}
As described in Section \ref{es_formalism}, the ES differential cross section is dependent on the values of $\sin^2\theta_W$, $g_V$, $g_A$, and a number of NSI parameters. Since
the elastic scattering process has an outgoing antineutrino, the visible energy, E$_{\mathrm{vis}}$, only provides a measure of the outgoing electron energy, rather than the antineutrino energy. The observed E$_{\mathrm{vis}}$ spectrum of the elastic
scattering events is a convolution of the IsoDAR electron antineutrino flux with the differential cross section 
given in Eq.~\ref{glgrxs}.  Experimentally, the elastic scattering events cannot be separated from the misidentified IBD beam and non-beam backgrounds described above.  Therefore, in measuring the parameters that enter the differential cross section, a fit is performed to the observed signal plus background events versus E$_{\mathrm{vis}}$.  The sensitivity of the measurement depends on the uncertainties associated with the predicted elastic scattering signal and the background events.

The misidentified IBD background, as described above, comes from the inefficiency in forming the delayed coincidence and has been estimated to be (0.25$\pm$0.02)\% of the total IBD sample.   The non-beam background is already being  measured by KamLAND~\cite{kamBoron}. It is assumed that more statistics will be accumulated during the construction of the IsoDAR source and that this data can be used directly with the only uncertainty being the statistical error associated with these beam-off measurements.

For the sensitivity estimates given here, the simulated signal and
background events are binned in 0.2~MeV E$_{\mathrm{vis}}$ bins from 3~MeV to
12~MeV.  A fit to the summed
distribution versus E$_{\mathrm{vis}}$ is then used to estimate the uncertainty achievable 
in a $\sin^2\theta_W$ measurement using a  $\Delta\chi^2 =
\chi^2(\mathrm{fit}) - \chi^2(\mathrm{best~fit})$ statistic.  For the sensitivity
estimate, one can assume that the best fit corresponds to the signal
with the input $\sin^2\theta_W^0 = 0.238$ and the input backgrounds.
Then the $\chi^2(\mathrm{best~fit}) = 0$ and $\Delta\chi^2 = \chi^2(\mathrm{fit})$.   

To be explicit,  the  form of the $\chi^2$ is written in the following way.
For the $ith$ bin of $E_{\mathrm{vis}}$,
we let $ES$ be the number of elastic scattering events at a given 
value of $\sin^2 \theta_W=s$.  Also, we define
$B^{\mathrm{on}}$ as the beam-on backgrounds (IBD mis-identification) and
$B^{\mathrm{off}}$ as the beam-off backgrounds for the $ith$ bin.    The
number of events in this bin will be 
\begin{equation}
N_{i}\left( s \right)  =ES_{i}\left( s\right)  +B^{\mathrm{off}}_{i}+B^{\mathrm{on}}_{i}
\end{equation}  

\noindent Then, if we abbreviate
$s_0=\sin^2\theta_W^0$ and $s_f=\sin^{2}\theta_{W}^{\mathrm{Fit}}$,
we can write:  
\begin{eqnarray}
&\chi^{2}&\left(  s_f \right)  = \nonumber \\
&\sum\limits_{i} &
\frac{\left(  N_{i}\left(  s_0 \right)  -\left(
N_{i}\left(  s_f \right)  +\alpha\ast ES_{i}\left(
s_f \right)  +\beta\ast  B^{on}_{i} \right)  \right)
^{2}}{\left(  N_{i}\left(  s_0 \right)
+B^{off}_{i}\right)  } \nonumber \\
&+&  \left(  \frac{\alpha}{\sigma_{\alpha}}\right)  ^{2}+\left(  \frac{\beta
}{\sigma_{\beta}}\right)^2  \label{Fit_chi2}~,
\end{eqnarray}
where the normalization uncertainties for the ES signal and IBD
misidentification background events are included using the pull
parameters $\alpha$ and $\beta$ respectively.  These parameters are
constrained by the measurements described above within the
uncertainties of  $\sigma_\alpha = 0.007$ and $\sigma_\beta = 0.02/0.25 = 0.08$.  Shape uncertainties for these backgrounds are negligibly small due to the precise measurement of the IBD event energy distribution.  The normalization and shape uncertainty for the beam-off background is taken from the assumed 4.5 years of beam-off data that will already have been accumulated by the KamLAND experiment.  The uncertainty for the beam-off background is included in the $\chi^2$ calculation by adding an additional error term to the statistical error in the denominator corresponding to this beam-off measurement ($B^{\mathrm{off}}_{i}$).

The beam-off background rates increase as the energy gets lower and the radius gets larger.  Optimization studies show that the fiducial volume restrictions with radial cuts from 5~m to 6~m yields similar $\sin^2\theta_W$ measurement sensitivity since the increase in backgrounds at higher radii counteracts the increased fiducial volume.  To minimize the sensitivity to these backgrounds, a radial cut of 5~m was chosen.   The differential cross section for antineutrino-electron scattering peaks towards low outgoing electron energy due to the energy carried away by the outgoing antineutrino.  Thus, a low E$_{\mathrm{vis}}$ cut will give the best $\sin^2\theta_W$ measurement sensitivity.  In order to avoid the many large backgrounds sources at low energy, a E$_{\mathrm{vis}}> 3$~MeV analysis cut is used.
\begin{table}[tb]
  \begin{center}
\begin{tabular}{r|cccc}
\hline
           &      Bkg factor & $\delta \sin^2\theta_W$ &$ \frac{\delta \sin^2\theta_W}{\sin^2\theta_W} $
           &  $\delta \sin^2\theta_W^\text{stat-only}$  \\
\hline
Rate +Shape &        1.0 &     0.0076 &      3.2\% &     0.0057 \\

Shape Only &        1.0 &     0.0543 &     22.8\% &     0.0395 \\

 Rate Only &        1.0 &     0.0077 &      3.2\% &     0.0058 \\

Rate +Shape &        0.5 &     0.0059 &      2.5\% &     0.0048 \\

Rate +Shape &        0.0 &     0.0040 &      1.7\% &     0.0037 \\
\hline
\end{tabular}      
\end{center}
\caption{ {Estimated $\sin^2\theta_W$ measurement sensitivity for various types of fits to the E$_{\mathrm{vis}}$ distribution.  The second column indicates the background reduction factor.} \label{results}}
\end{table}
\begin{figure}[tb ]
	\begin{center}		
		\includegraphics[width=0.49\textwidth]{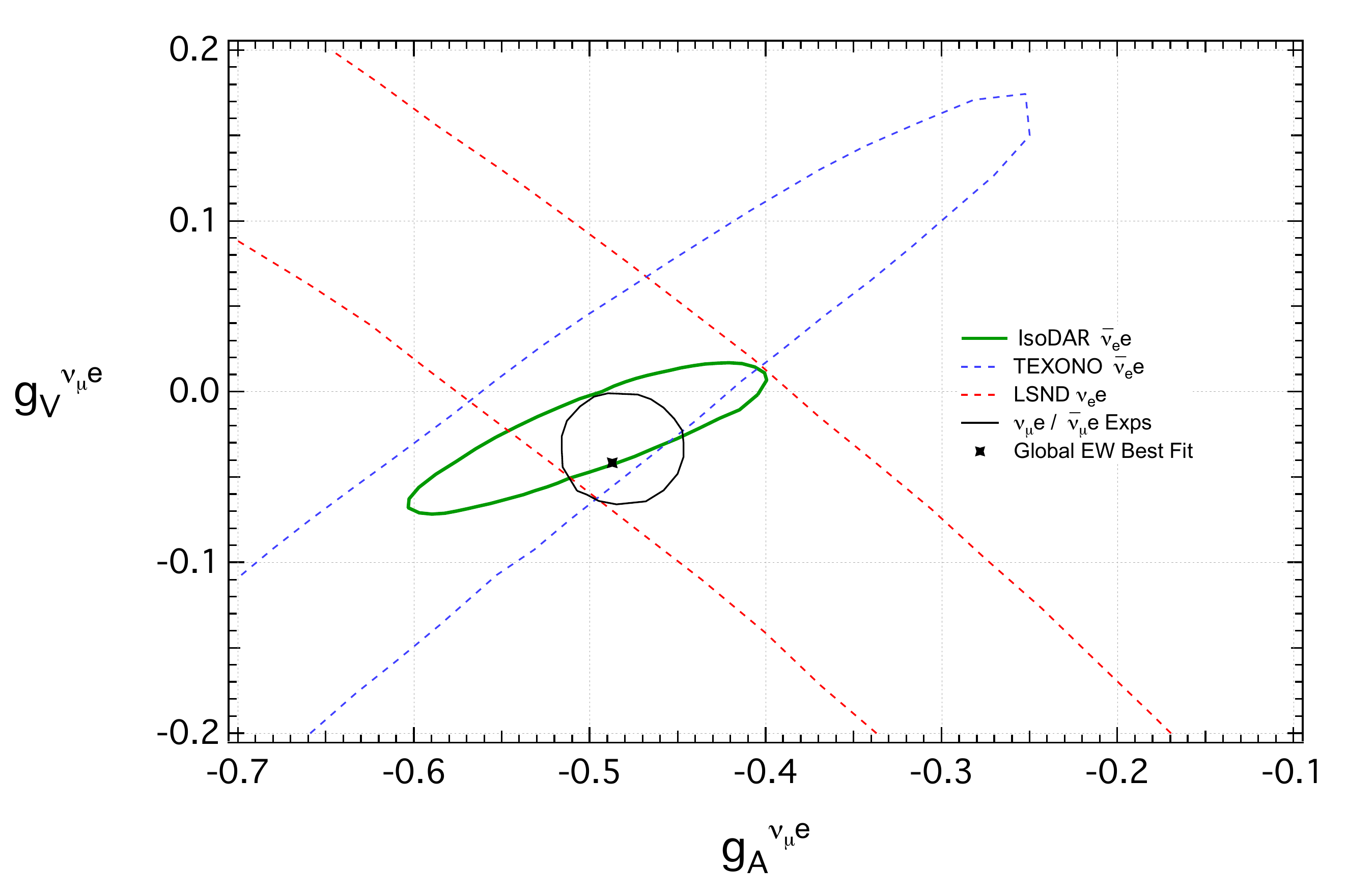}
	\end{center}
	\caption{\label{gV_gA} IsoDAR's sensitivity to $g_V$ and $g_A$ along with allowed regions from other neutrino scattering experiments and the electroweak global best fit point taken from Ref.~\cite{PDG2}. The IsoDAR, LSND, and TEXONO contours are all at $1\sigma$ and are all plotted in terms of $g_{V,A}^{\nu_\mu e}= g_{V,A}^{\nu_e e}-1$ to compare with $\nu_\mu$ scattering data. The $\nu_\mu e/\bar{\nu}_\mu e$ contour is at 90\% C.L.}
	\end{figure}
\begin{figure}[tbh ]
	\begin{center}		
		\includegraphics[width=0.49\textwidth]{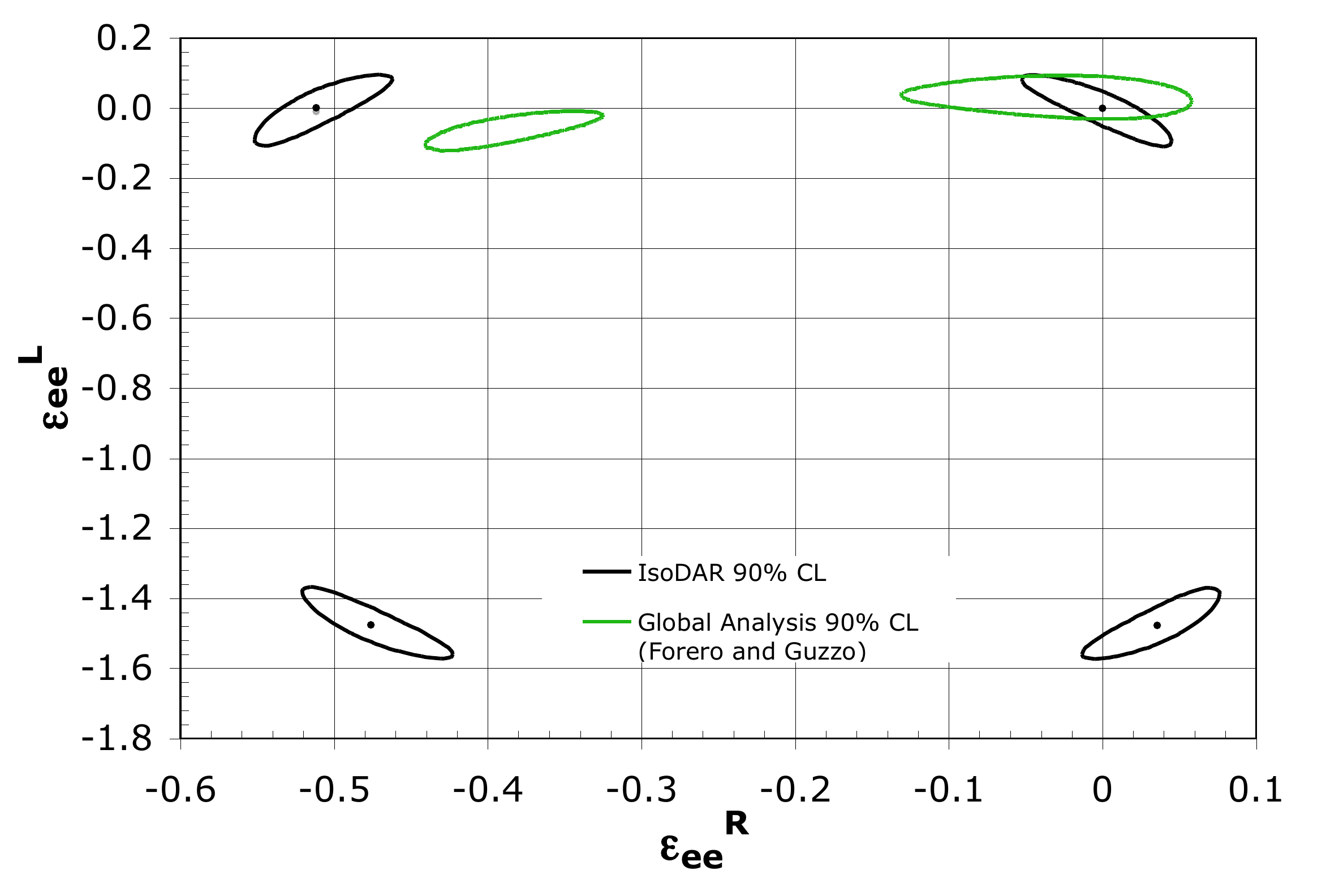}
		\\
		\includegraphics[width=0.49\textwidth]{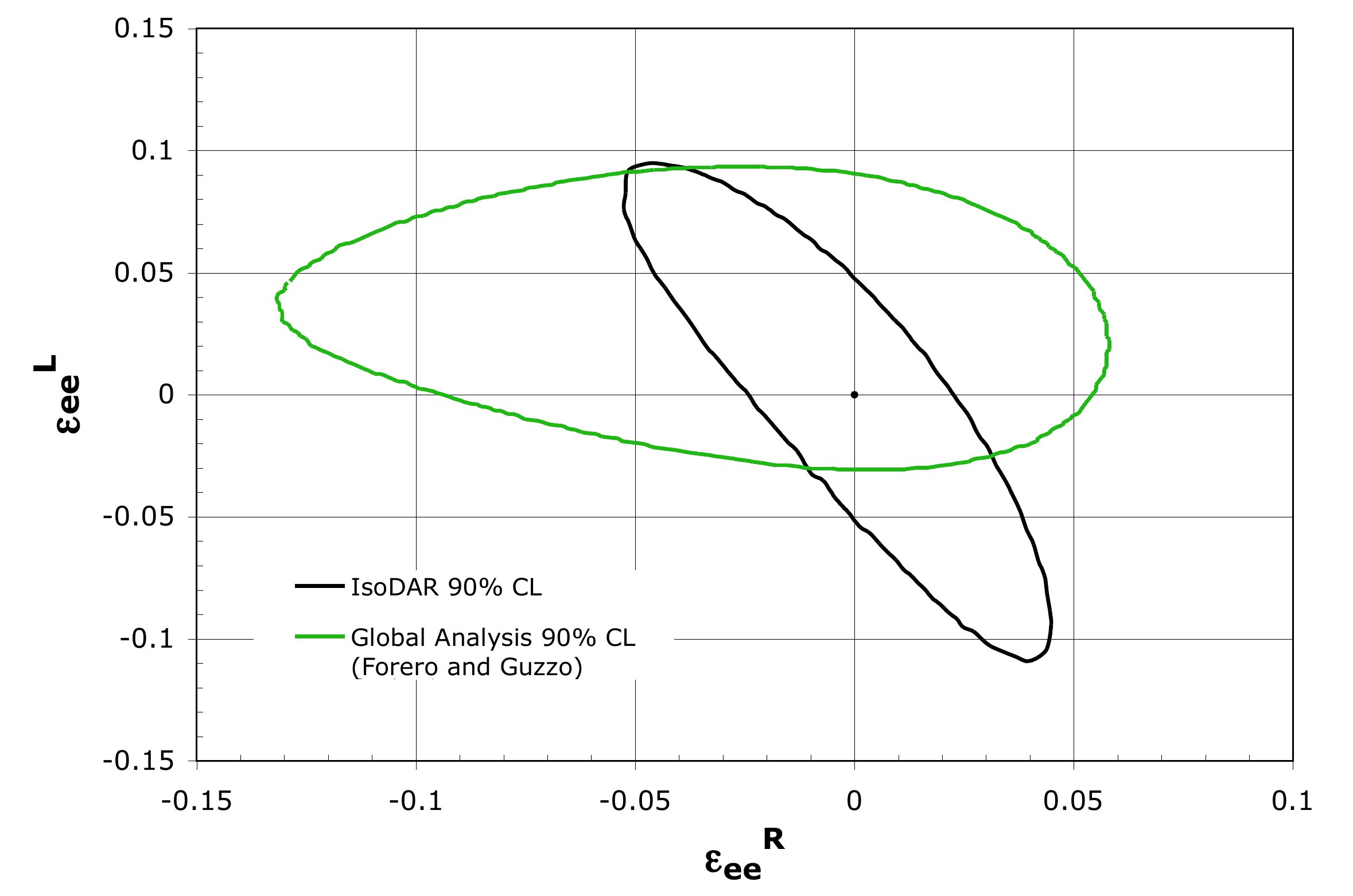}
	\end{center}
	\caption{\label{NSI} (Top) IsoDAR's sensitivity to $\epsilon_{e e}^{e L}$ and $\epsilon_{e e}^{e R}$. The current global allowed region, based on Ref.\cite{NSI_forero_guzzo} is also shown. (Bottom) A zoomed-in version of the top plot, emphasizing the region near $\epsilon_{ee}^{e  L}$ and $\epsilon_{ee}^{e  R}$ $\sim0$ is shown.}
\end{figure}

With the cuts previously described and with the assumptions listed in Table~\ref{assumptions}, the total numbers of elastic scattering and background events are given in Table~\ref{events}.  Fits to the E$_{\mathrm{vis}}$ distribution of the event sum, using the $\chi^2$ function given in Eq.~\ref{Fit_chi2}, yields the results shown in Table~\ref{results}. The results are given for a combined fit of the rate and E$_{\mathrm{vis}}$ shape along with each separately.  From these results, it is clear that this measurement is mainly dependent on the sensitivity of the rate to changes in $\sin^2\theta_W$ and is dominated by statistical uncertainty.  The slope, $d\sin^2\theta_W / d N = 7.4\times 10^{-5}$, when combined with the total event rate of 6158.8 implies a statistical uncertainty on $\sin^2\theta_W$ of 0.0058. Backgrounds could be reduced further using more advanced analysis techniques. For example, if the directionality of the incoming antineutrino could be reconstructed\cite{christoph}, the ES events could be effectively separated from isotropic backgrounds.  Results are also shown for the case where the background is reduced by 50\% or eliminated. 

In addition we can treat Eq.~\ref{Fit_chi2} as a function of $g_V$ and $g_A$ and perform a two-parameter fit.  The $1\sigma$ contour for this fit is shown overlaid on data from other neutrino electron scattering experiments in Fig.~\ref{gV_gA}.  The charge current contribution has been removed from the $\nu_e e$ and $\bar{\nu}_e e$ scattering data by plotting the contours in terms of $g_{V,A}^{\nu_\mu e}= g_{V,A}^{\nu_e e}-1$ in order to more easily compare with $\nu_\mu e$ and $\bar{\nu}_\mu e$ scattering data. IsoDAR significantly constrains the global allowed region for the weak couplings derived from $\nu_e e$ and $\bar{\nu}_e e$ scattering data and can test their consistency with the weak couplings derived from $\nu_\mu e$/$\bar{\nu}_\mu e$ scattering.

Finally, using the assumptions listed in Table~\ref{assumptions} as well as the background and systematics previously described, we can also estimate IsoDAR's sensitivity to the NSI parameters $\epsilon_{ee}^{e
  L}$ and $\epsilon_{ee}^{e
  R}$. The results are shown in Figure~\ref{NSI} along with the current global allowed region~\cite{NSI_forero_guzzo}. The four-fold degeneracy arises due to the cross section's dependence on the square of the NSI parameters. The degeneracy can potentially be broken with the aid of the world's data, especially measurements involving neutrinos.  In the region around $\epsilon_{ee}^{e  L}$ and $\epsilon_{ee}^{e  R}\sim0$, the IsoDAR 90\% confidence interval would significantly improve on the allowed global fit region for $\epsilon_{ee}^{e L R}$.
  
 \section{Conclusions}
A  pure, low energy $\bar \nu_e$ source produced through
$^8$Li $\beta$-decay by the IsoDAR source, in combination with the KamLAND detector, can produce a sample of more than 2400 ES events in 5~years of running.    This large sample can be used for a precise measurement of $\sin^2 \theta_W$ and and to test the consistency of weak couplings measured in $\nu_e e$ and $\nu_\mu e$ scattering data. Using a measurement strategy inspired by~\cite{ConradLinkShaevitz} and a background model based on~\cite{kamBoron} we perform a $\chi^2$ analysis on the differential ES cross section and derive a 3.2\% measurement sensitivity on $\sin^2\theta_W$. This would be the most precise determination of $\sin^2\theta_W$ from $\nu_e e$ or $\bar{\nu}_e e$ scattering data.  The IsoDAR ES sample can also serve as a sensitive probe of nonstandard neutrino interactions.  A two-parameter $\chi^2$ fit to the two non-universal NSI parameters $\epsilon_{ee}^{eLR}$ would allow for a sensitive search for new physics beyond the current global fits.

\begin{center}
{ \textbf{Acknowledgments}}
\end{center}
The authors thank Will Loinaz for useful discussions.  The authors also thank the attendees of the 2012 Erice International School of Subnuclear Physics Workshop for valuable discussions. Support for this workshop was provided through the Majorana Centre from the INFN Eloisatron Project directed by Prof. Antonino Zichichi.  The authors thank National Science Foundation for support. LW is supported by funds from UCLA.  
                                                             
\bibliography{anue_v2}

\end{document}